\documentclass[12pt,preprint]{aastex}   
\def\paren#1{\left( #1 \right)}

\def\Mesz{M\'esz\'aros~} 
\begin{document}
\title{The Onset of Gamma-Ray Burst Afterglow}
\author{Shiho Kobayashi\altaffilmark{1} and
        Bing Zhang \altaffilmark{2}} 
\altaffiltext{1}{Astrophysics Research Institute, Liverpool 
                 John Moores University, Birkenhead CH41 1LD, UK}
\altaffiltext{2}{Department of Physics,
                 University of Nevada, Las Vegas,
		 NV 89154, USA}
\begin{abstract}
 We discuss the reference time $t_0$ of afterglow light curves in the
 context of the standard internal-external shock model. The decay index
 of early afterglow is very sensitive to the reference time one
 chooses. In order to understand the nature of early afterglow, it is
 essential to take a correct reference time. Our simple analytic model
 provides a framework to understand special relativistic effects
 involved in early afterglow phase. We evaluate light curves of reverse
 shock emission as well as those of forward shock emission, based on
 full hydrodynamic calculations. We show that the reference time does
 not shift significantly even in the thick shell case. For external
 shock emission components, measuring times from the beginning of the
 prompt emission is a good approximation and it does not cause an early
 steep decay. In the thin shell case, the energy transfer time from
 fireball ejecta to ambient medium typically extends to thousands of
 seconds. This might be related to the shallow decay phases observed in
 early X-ray afterglow at least for some bursts. 
\end{abstract}
\section{Introduction}
It is well known that gamma-ray burst (GRB) afterglow decays as a
power-law $L\propto(t-t_0)^{-\alpha}$.  The temporal decay index $\alpha$,
together with the spectral index, provides us precious information about
GRB jets and their environment. In the pre-Swift era afterglow
observations start typically a few hours after a burst. In such a late
phase, the decay index is insensitive to the choice of the reference
time $t_0$, and GRB trigger time is often used in afterglow modelings.

The multi-wavelength observatory Swift was launched in Nov 2004. Thanks
to its fast pointing capabilities Swift is disclosing the early
afterglow phase. One of unexpected finds by Swift is that early X-ray
afterglows show a canonical behavior, where light curves include three
components: (1) a steep decay component, (2) a shallow decay
component and (3) a ``normal'' decay component. On top of this canonical
behavior, many events have superimposed X-ray flares (Zhang et al. 2006a;
Nousek et al. 2006; Chincarini et al 2005; O'Brien et al. 2006). The
transition from the early steep decay to the shallow decay typically
occurs at several hundred seconds, and the timescale is comparable to
the duration of rather long GRBs. When discussing the early
afterglow and its connection to the prompt emission component, the decay
index is very sensitive to the reference time $t_0$ one
chooses. Correctly choosing $t_0$ is therefore essential to derive the
right index as well as to interpret each component in the canonical
light curve (Piro et al. 2005; Tagliaferri et al. 2005; Quimby et al. 
2006). 

Tagliaferri et al. (2005) investigated the first two bursts GRB 050126
and GRB 050219a which have an X-ray light curve well sampled by the
X-Ray Telescope on board Swift. They sought for a possible delay of 
the afterglow onset by fitting the early X-ray light curves (the components
1 and 2 which we have discussed above) with a single power-law model. In
both cases, the decaying light  curves can be fitted if the 
onset of the afterglow is shifted to $t_0 \sim 100$ s after the burst
trigger with a single power-law. However, while in the case of GRB
050126 the light curve does not allow us to clearly state whether a 
broken power-law modeling is better than a single power-law model, 
for GRB 050219a a broken power-law definitively provides a better fit. 

In the standard GRB model the time shift between the GRB trigger and the
reference time $t_0$ is expected to be ``small''. The early steep 
decay should not be an artifact due to a wrong choice of $t_0$. Lazzati and
Begelman (2006) studied forward shock emission, based on a simple energy
injection model. Their numerical light curves show that measuring times
from the beginning of the prompt phase is a good approximation. The
early steepening and X-ray flares are likely to be  produced by another
mechanism  (e.g. internal shocks; Burrows et al. 2005; Falcone et
al. 2006; Zhang et al. 2006a; Nousek et al. 2006; Ioka, Kobayashi \&
Zhang 2005; Fan \& Wei 2005).  
Recently long-lasting soft emission is reported in a short burst GRB
050724 (Barthelmy et al. 2005). Such a soft component was hinted at in
the sum of multiple short BATSTE GRBs, and it might be the onset of
short burst afterglow (Lazzati, Ramirez-Ruiz \& Ghisellini 2001). It is
therefore of interest to quantitatively examine the $t_0$ issue.

In this paper, we study the physics and timescales involved in early
afterglow stage, and give more direct and clear arguments for the 
afterglow slopes. In \S 2 we study a simple analytic model. In \S 3 we
evaluate light curves of reverse shock emission as well as that of
forward shock emission, based on full hydrodynamic calculations. 
In \S 4 we address how inhomogeneity of a fireball affects an early
afterglow light curve. Conclusions and discussion are given in \S 5.      

\section{The Reference Time $t_0$}

\begin{figure}
\plotone{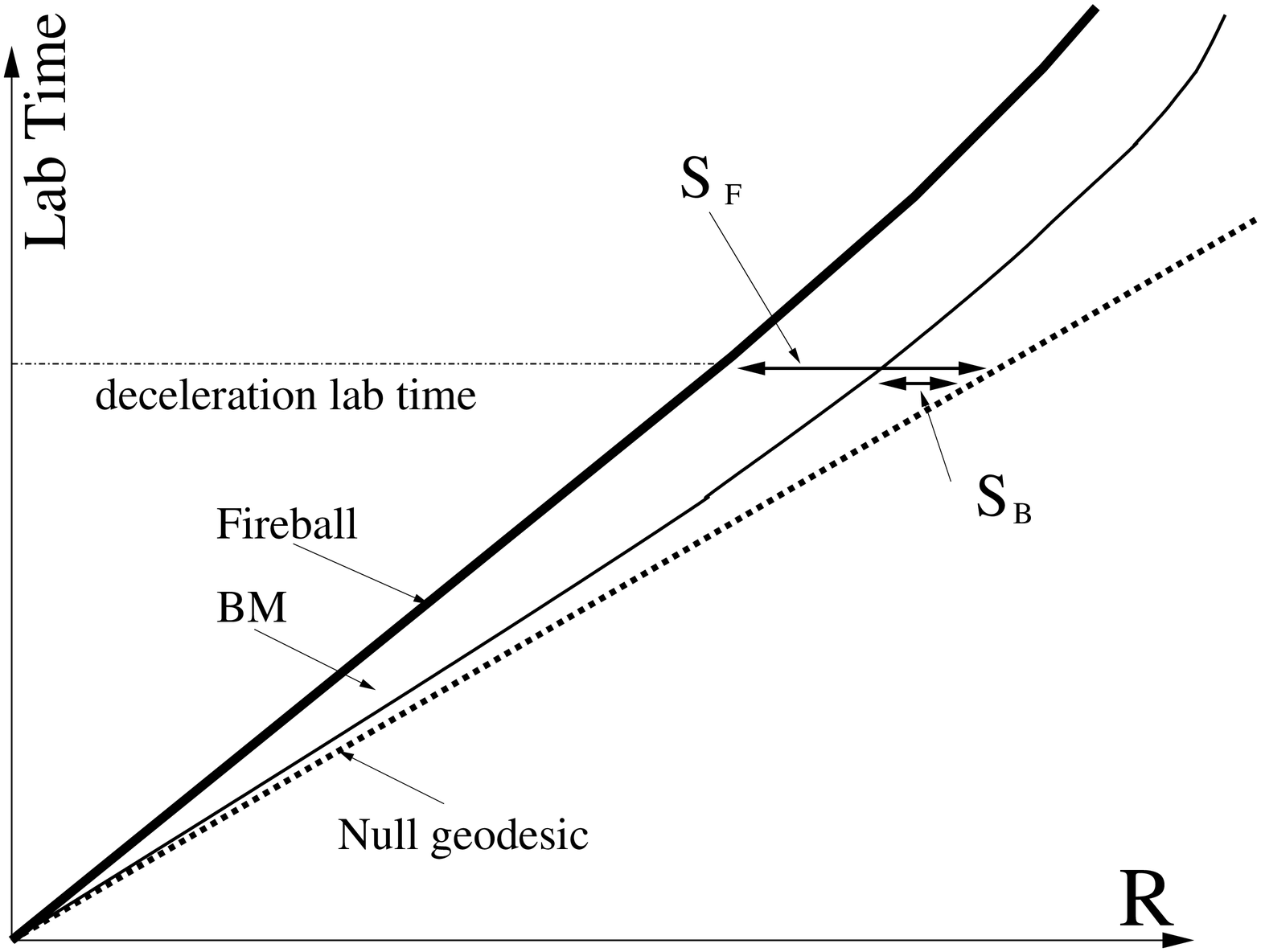}
\caption{Spacetime diagram: 
  If the evolution of a fireball satisfies the BM blast wave scaling
  $\Gamma \propto R^{-3/2}$  from the beginning, the thin solid 
  curve gives the trajectory. However, 
  the fireball initially coasts with a finite Lorentz factor. The evolution is
  described by the BM solution only after it begins to be decelerated. 
  At the deceleration lab time the fireball
  (thick solid) gets behind the reference BM blast wave (thin solid), and the 
  trajectories approaches each other at a later time.
  The observed time is given by the delay of photons from an emitter at
  a lab time with respect to the gamma-ray front (null geodesic; dashed
  line).  Null geodesics are trajectories of photons. Physical
  trajectories should be steeper or parallel to this minimal  
  slope line. A trajectory of an object at rest is vertical.
  \label{fig1}}
 \end{figure}

Let $R$ be a radius of a forward shock expanding with a Lorentz factor
$\Gamma \gg 1$ into homogeneous ambient medium. Since the shock is
moving toward us at almost the speed of light $c$, the difference of the
the observed times between a photon emitted at $R$ and another emitted
at $R+dR$ is $dt\sim dR/2c\Gamma^2$. Although the origin of observed
time is arbitrary, a natural definition of observed time is given by the
delay of photons emitted from a shock front at a lab time $\hat{t}$ 
with respect to the photons emitted from the ``explosion'' at $R=0$
and $\hat{t}=0$. The dashed line in figure \ref{fig1} depicts the
trajectory of the photon from the explosion. 
In this paper, $\hat{t}$ and $t$ denote the lab and observed time since 
the explosion, respectively. Distance, velocity and the corresponding
Lorentz factors are measured in the lab frame. Thermodynamic quantities
(pressure and density) are measured in the local fluid frame.
The observed time is given by
\begin{equation}
t=\int_0^R \frac{dR}{2c\Gamma^2}.
\label{eq:obstime}
\end{equation}
The shock radius is almost proportional to the lab time,
$\hat{t}\sim R/c$.
Considering that internal shocks occur and produce gamma-rays at radii
much smaller than the deceleration radius of the fireball, the prompt
gamma-rays associated with the outermost element of the fireball
should propagate practically on the dashed line in figure 
\ref{fig1}. The GRB trigger almost coincides with the explosion
$t=0$. The difference is small, and order of the variability timescale
of the prompt emission, because the variability timescale directly reflects
the inhomogeneity scale in a fireball (Kobayashi, Piran \& Sari 1997). 
The dashed line in figure \ref{fig1} will be called the 
gamma-ray front in the following sections.

A fireball with an initial Lorentz factor $\Gamma_0$ decelerates when  
it collects a large volume of ambient material with a mass density 
$\rho_1$.
Equalizing energy of the shocked ambient material $4\pi
R^3\rho_1c^2 \Gamma_0^2/3$ and fireball energy $E$, we obtain the
deceleration radius $R_d=l/\Gamma_0^{2/3}$ where 
$l=(3E/4\pi\rho_1c^2)^{1/3}$ is the Sedov length.  Since the fireball
density decreases as it expands, there is a possibility that a reverse
shock evolves from Newtonian to relativistic during the propagation. A
relativistic reverse shock reduces considerably the Lorentz factor of
the fireball material which it crosses. In such a case, the energy of
the shocked ambient medium is still negligible at $R=l/\Gamma_0^{3/2}$,
because we have assumed $\Gamma \sim \Gamma_0$ at the deceleration to
estimate $R_d$. 
The reverse shock crosses the fireball shell at $R=l^{3/4}\Delta^{1/4}$ 
and all the fireball material decelerates where $\Delta$ is a fireball 
shell width (see Sari \& Piran 1995  and Kobayashi, Piran \& Sari 1999
for the details). In summary, the evolution of fireballs are classified
into two cases depending on the value of $\Delta$ relative to a critical
value $\Delta_c=l/\Gamma_0^{8/3}$ (Sari \& Piran 1995). If $\Delta$ is smaller
than the critical value, it is called the thin shell case
\footnote{In the thin shell case, the deceleration observed time is
given roughly by the critical width $\sim \Delta_c/c= R_d/c\Gamma_0^2$,
and it is longer than the duration of the prompt emission 
$\Delta/c$ (Sari 1997). The deceleration time $\Delta_c/c$ approaches
$\Delta/c$ if we take a larger $\Gamma_0$, and around $\Gamma_0 \sim
\Gamma_c=(l/\Delta)^{3/8}$ (or equivalently $\Delta \sim \Delta_c$ ) a
reverse shock becomes relativistic during the propagation. The Lorentz
factor of the shocked material at the crossing time becomes independent 
from the initial value, and it is given by $\Gamma_c$.
The deceleration observed time is about $R_d/c \Gamma_c^2=\Delta/c$.}. 
The reverse shock is always in the
Newtonian regime, and it is too weak to slow down the fireball
effectively. The deceleration radius is $R_{d}=l/\Gamma_0^{2/3}$. If
$\Delta > l/\Gamma_0^{8/3}$ (the thick shell case), we define the
deceleration radius as the shock crossing radius
$R_d=l^{3/4}\Delta^{1/4}$. The deceleration lab time is given by 
$\hat{t}_d \sim R_d/c$.   

If the Blandford-McKee (BM) blast wave scaling $\Gamma \propto
R^{-3/2}$  (Blandford and McKee 1977) was valid for the whole fireball
evolution (the thin solid line in figure \ref{fig1}), integrating
eq. (\ref{eq:obstime}), we obtain $R\propto t^{1/4}$. Since the spectral
characteristics of forward shock synchrotron emission are 
given by products of $t$, $R(t)$, $\Gamma(t)$ and constant parameters
(Sari, Piran \& Narayan 1998), the light curve should be described by a
power-law with the reference time $t_0=0$. However, the BM blast wave
scaling is applicable only after the fireball is decelerated
(e.g. Kobayashi, Piran \& Sari 1999). At earlier times the fireball (the  
thick solid line) is in the coasting phase, and it is slower than
evaluated from the BM blast wave scaling. 
The delay of photons from the shock front at the deceleration
lab time is larger by  $\delta t=(S_F-S_B)/c$ than in the case that the
BM blast wave scaling is applicable to the whole evolution (see figure
\ref{fig1}) where $S_B$ and $S_F$ are the separations at the
deceleration lab time between the gamma-ray front (dashed line) and the
reference BM blast wave (the thin solid line), and between the gamma-ray
front and the fireball forward shock (the thick solid line), respectively. 
The reference time for the afterglow modeling is given by 
\begin{equation}
t_0\equiv\frac{S_F-S_B}{c},
\end{equation}

\begin{figure}
\plotone{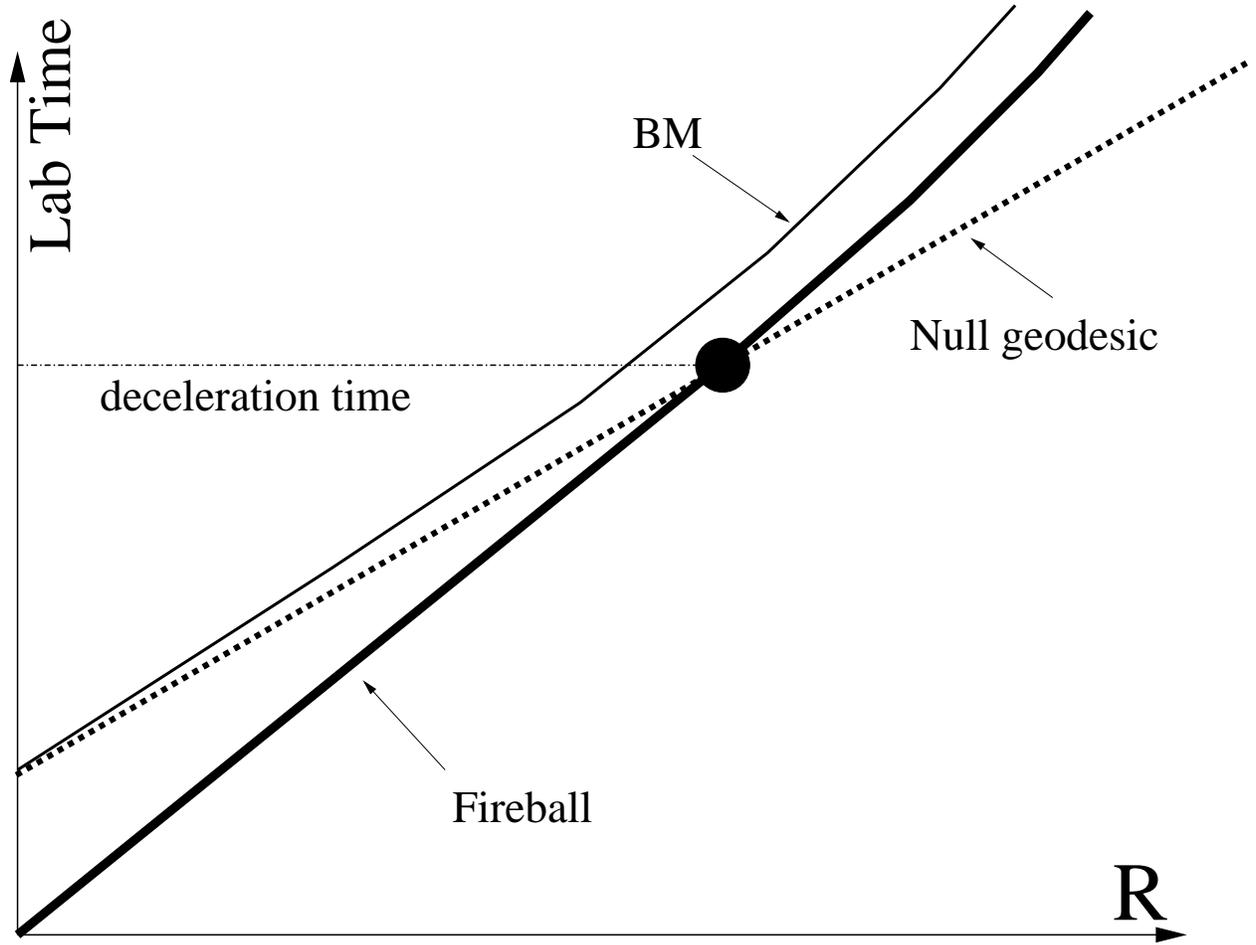}
\caption{wrong choice of $t_0$ : Fireball Evolution
  (thick solid curve), a shifted null geodesic (dashed line) and its 
  reference BM blast wave (thin solid).
 \label{fig2}}
 \end{figure}

In the thick shell case, a forward shock keeps being energized for a
longer time, and the deceleration phase starts at a later time. It still
overestimates the shift of the reference time if $t_0$ is set at the end
of the energy injection or equivalently at the peak time of afterglow.
In figure \ref{fig2} the dot indicates the point at which the energy
injection stops and the fireball turns into a BM solution. Measuring
times from the peak of afterglow corresponds to defining a reference
null geodesic (the dashed line) as it goes through the dot. 
Imagine the reference BM blast wave (the thin solid line) associated
with the new reference null geodesic
\footnote{BM blast wave lines on the spacetime diagram are determined by two
parameters, an explosion time (i.e. a null geodesic on the diagram) and
the Sedov length. The latter is evaluated with the total fireball energy
if there is energy injection before the deceleration time.}.
The decay of the emission from this reference blast wave should be
characterized by a power-law using the new reference time (measuring 
times from the peak). The evolution of a real fireball (the thick solid
line) is also described by a BM blast wave after the deceleration.
However, it is a different solution (a different curve on the diagram).
Applying the new reference time to the afterglow modeling leads to a
wrong estimate of the decay index especially in the early phase. The
decay index should become shallower than the real value.  

Using eq.(\ref{eq:obstime}) and the evolution of a BM blast wave $\Gamma
\sim (R/l)^{-3/2}$, we obtain $S_B=R_d/8\Gamma_d^2$ where $\Gamma_d$ is
the Lorentz factor at the deceleration radius $R_d$. Since the Lorentz
factor of a fireball is constant $\Gamma=\Gamma_0$ at $R<R_d$, 
we get $S_F=4 S_B$ for the thin shell. In the thick shell case,
a reverse shock becomes relativistic before it crosses the fireball
shell, and it begins to reduce considerably the Lorentz factor of the shell's 
matter which it crosses. The Lorentz factor of the forward shock 
is also significantly reduced as $\Gamma \sim (l^3/\Delta R^2)^{1/4}$ during the
reverse shock crossing (Sari \& Piran 1995; Kobayashi, Piran \& Sari
1999), and  we get $S_F=2 S_B$ for the thick shell.
The reference time is given by
\begin{equation}
t_0=t_d\paren{1-\frac{S_B}{S_F}}=
\left\{
\begin{array}{@{\,}ll}
3t_d/4  &  \mbox{the thin  shell case}   \\
t_d/2   &  \mbox{the thick shell case}   
\end{array}
\right. 
\label{eq:t0}
\end{equation} 
where $t_d=S_F/c$ is the deceleration observed time. 

We have considered a simple broken power-law model for the evolution of
a fireball Lorentz factor, and we obtained eq. (\ref{eq:t0}). In reality the  
fireball should decelerate gradually around the deceleration radius. 
An artificial early steepening could happen  only when $S_B \ll S_F$,
and in such a case $t_0$ should be set at $t_d$ (the afterglow peak time) to
get the correct $\alpha$. However, in
both of the thin shell and thick shell cases, we have found that $S_B$
and $S_F$ are comparable. Therefore, measuring times from the beginning
of the prompt phase $t_0=0$ should not induce an overestimate of
afterglow decay right after the peak.

Eq. (\ref{eq:obstime}) gives the delay of photons emitted from a point
on the shock front on the line of sight, while most photons suffer
longer delays, since they are emitted from a shocked region of finite
thickness behind the shock, and from positions off the line of sight.
Although these effects could make both separations $S_B$ and $S_F$ 
larger by a factor of a few (Waxman 1997), they are still comparable and 
our arguments are valid.

\section{Numerical Model}
We employ a spherical relativistic Lagrangian code based on the Godunov
method with an exact Riemann solver to evaluate the hydrodynamic
evolution of a relativistic fireball (Kobayashi et al. 1999; Kobayashi
\& Sari 2000). Using the Einstein summation convention the equations 
describing the motion of a relativistic fluid are given by the five 
conservation laws
\begin{equation}
\partial(\rho u^i)/\partial x^i=0, \ \ \partial T^{ik}/\partial x^k=0,
\end{equation}
where $u^i$ is four velocity and $T^{ik}$ is the stress-energy tensor 
($i,k=0,...,3$), which for a perfect fluid can be written as $T^{ik}=w
u^iu^k-pg^{ik}$. Here, $g^{ik}$ is the metric tensor, $p$
the fluid pressure, and $w=e+p$ the heat function per unit volume.
Shocked material is extremely hot and pressure is related to
internal energy $e$ and mass density $\rho$ as $p=(e-\rho c^2)/3$.
In the Godunov scheme, conservative variables are considered as
piecewise constants over the mesh cells at each time step and 
the time evolution is determined by the solution of the Riemann problem
(shock tube) at the inter-cell boundaries (e.g. Mart\'i \& M\"uller 1999
and references therein).

Because of the relativistic beaming effect, the radiation 
from a jet before the jet break can be described by a spherical model with
an isotropic energy. The initial configuration for our simulation is a
static uniform fireball surrounded by uniform cold ambient
material (ISM). It is determined by four parameters: an isotropic energy 
$E$, a dimensionless entropy $\Gamma_0$, an initial radius $R_0$ and the ISM
mass density $\rho_1$. $E$ and $\rho_1$ always appear as the ration
$E/\rho_1$ in the hydrodynamics computation, the system is actually
determined by three parameters, the initial radius $R_0$, the entropy
$\Gamma_0$ and the Sedov length $l$. For convenience, we set the initial
lab time as $R_0/c$ rather than zero in numerical calculations, and the
observed time is determined  by the delay of photons from an emitter
with respect to the photons emitted from the initial fireball surface at
$R=R_0$ and $\hat{t}=R_0/c$. Since the fireball immediately accelerates
to a relativistic velocity, this is practically equivalent to 
the observed time used in figure \ref{fig1}. 

We carry out numerical calculations for the total number 
\footnote{We have reevaluated the light curves Figs \ref{thinLC} and
\ref{ThickLC} for $N=5400$.  In the log-log space, the resulting light
curves are almost identical. The differences are less than several
percent. When we remove numerical 
oscillations, they are in a few percent agreement.}
of mesh cells $N=540$. 
A third of the cells are for the fireball, while the other
cells describe the ambient medium within $R_{max}= 10^{18}$cm. Although
the initial configuration is $\sim 55$ cells per decade, the nature of
Lagrangian method gives a much higher resolution for shocks
which sweeps cells and compresses them by a factor of $\sim \Gamma^2$. 
A reflection boundary and free boundary condition are imposed at the
center $R=0$ and at $R_{max}$, respectively.
We consider two cases -- the thin and thick shell cases.  

\subsection{The Thin Shell Case}
We first consider the thin shell case: $E=10^{53}$ ergs, $\Gamma_0=100$, 
$R_0=3\times 10^{11}$cm, and $\rho_1=1$ $m_p$ cm$^{-3}$ where 
$m_p$ is the proton mass. 
Initially, as the fireball expands into a  surrounding medium, a narrow
shell with a radial width $\Delta \sim R_0$ is  
formed. The Lorentz factor of the shell increases linearly with the 
radius during the free acceleration stage. Then, the fireball shell
uses up all its internal energy, and it coasts with the Lorentz factor 
of $\Gamma_0$. The coasting ends once the ISM begins to influence the
shell. After the deceleration radius $R_d$, the profile of the shocked
ISM medium begins to approach the BM solution.
The evolution of a fireball is fully discussed by Kobayashi et al. (1999).    

\begin{figure}
\plotone{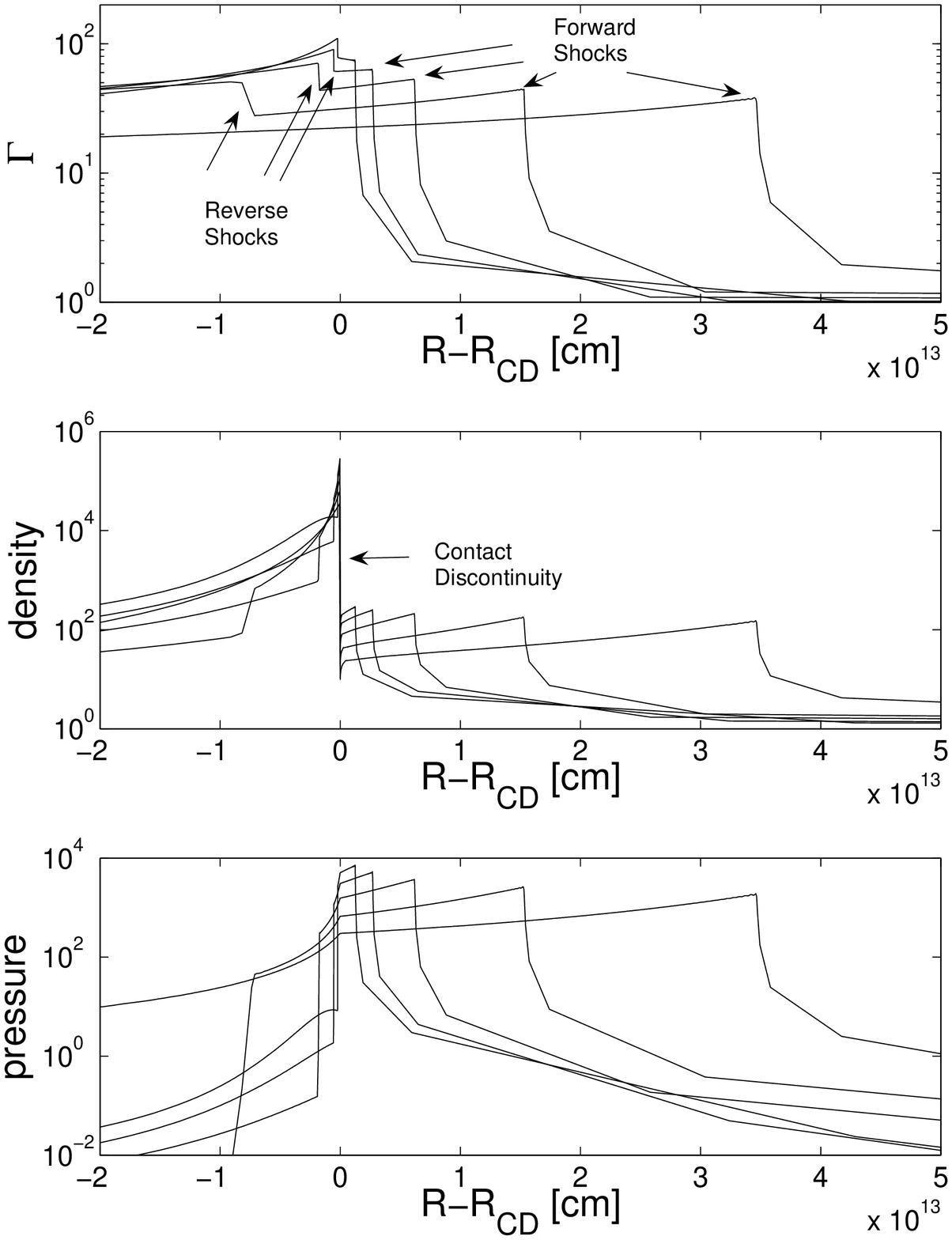}
\caption{Thin Shell Case: Profiles of $\Gamma$, $\rho$ and $p$ at different lab times
  $\hat{t}=0.62 R_d/c$, $0.86 R_d/c$, $1.14R_d/c$, $1.45 R_d/c$ and
  $1.74R_d/c$. The $x-$axis is the distance from the
  contact discontinuity (CD). Photons emitted from the discontinuity 
  at the lab times are observed at $t=100, 200, 400, 840$ and 
  1650 s, respectively. The gamma-ray front
  is located at ($R-R_{CD})=ct$. 
 \label{profNRS}}
 \end{figure}

\begin{figure}
\plotone{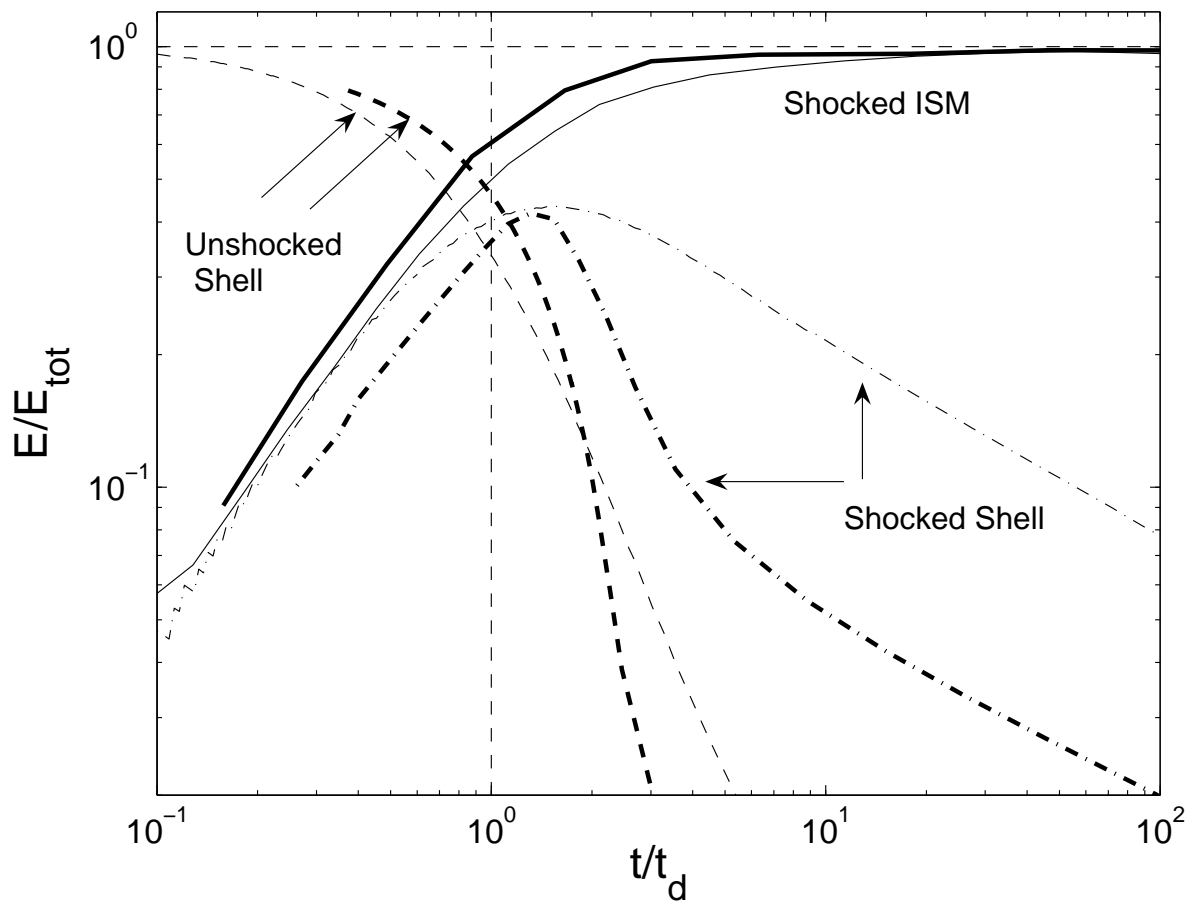}
\caption{Energy transfer from the fireball to the shocked 
regions. the thin shell case (thin lines) and the thick shell case
(thick lines). The sum of the kinetic energy and the thermal energies  
in the unshocked shell (dashed lines), in the reverse shocked 
shell (dashed dotted lines) and in the forward shocked ISM
(solid lines). All the energies are normalized by the explosion
energies. The observed time is normalized by the deceleration time $t_d$.
 \label{energy}}
 \end{figure}

The interaction between the shell and the ISM is described by
two shocks: a forward shock propagating into the ISM and a
reverse shock propagating into the shell. Figure \ref{profNRS} shows 
the propagation of the shocks. Initially, the unshocked 
fireball shell has all the energy of the system. As the shell expands, 
the reverse shock decelerates the ejecta while the forward shock
accelerates the ISM. The energy is transfered from the
unshocked shell to the ISM via the shocks, finally the shocked ISM
carries essentially all the energy of the system. In the intermediate
stage, around the deceleration time
$t_d=(3E/32\pi\rho_1c^5 \Gamma_0^8)^{1/3}\sim 195$ sec,
the shocked shell has comparable energy to the shocked ISM. The
evolution of the  
energies in three regions, inside of the reverse shock (unshocked
shell), between the reverse  shock and the contact discontinuity
(shocked shell), and between the contact discontinuity and the forward
shock (shocked ISM) are shown in Figure \ref{energy}. The observed times
of photons from the reverse shock front, the contact discontinuity and the 
forward shock front are used to describe the evolution of the energies. 

Even after the deceleration time, the reverse shocked shell (the thin
dashed-dotted line) carries a significant fraction of the system energy
for a long time, $20\%$ at $t\sim 10 t_d$ and $10\%$ at $t\sim 45 t_d$.
In the thin shell case, the reverse shock is Newtonian or sub-relativistic
in the frame of the unshocked shell (the deceleration by the reverse
shock is not significant as we can see in the top panel of figure
\ref{profNRS}). It does not heat the shocked region well. The reverse 
shocked region is already cold at the shock 
crossing time (the deceleration time), and the shocked shell carries the energy
mainly in the form of the kinetic energy $E \propto \Gamma$. Assuming a
power-law decay $\Gamma \propto R^{-g}$, we obtain $E \propto \Gamma
\propto t^{-g/(1+2g)} \sim t^{-0.4}$ for $g=2.2$ (Kobayashi \& Sari
2000). 

\begin{figure}
\plotone{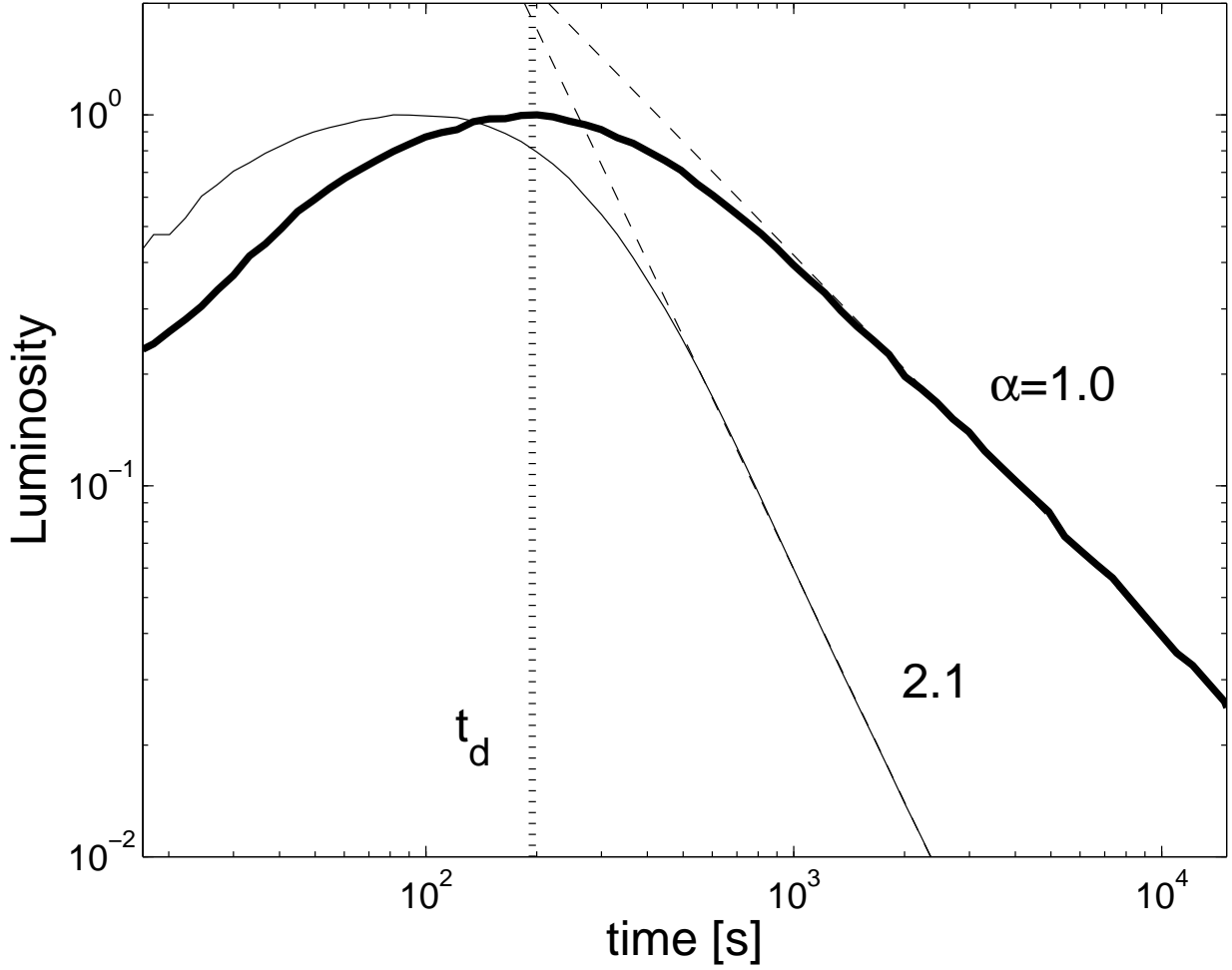}
\caption{Afterglow Light Curve: the Thin Shell Case.
$E=10^{53}$ ergs, $\Gamma_0=100$, $R_0=3\times 10^{11}$cm, 
  $\rho_1=1$ $m_p$ cm$^{-3}$ and $\hat{p}=2.2$. 
 Forward shock emission in X-ray band (thick curve) 
 and reverse shock emission in optical band (thin curve).
 Power-law fits (dashed lines): 
 forward shock $\alpha=1.0$ and reverse shock $\alpha=2.1$.
 The light curves are normalized at the peaks.
 \label{thinLC}}
 \end{figure}

This slow energy transfer should lead to the round-off of an afterglow
peak. To evaluate the afterglow light curve, we consider here a simple
case in which the energy of the magnetic field remains a constant
fraction of the internal energy $B^2\propto p$. The electron random
Lorentz factor evolves as $\gamma_m \propto p/\rho$ after the shock heating
where $p$ and $\rho$ are the pressure and density of a fluid element.
The typical synchrotron frequency in the observer frame is 
$\nu_m \propto \Gamma \gamma_m^2B$. Since a relativistic shock totally
ionizes material which it crosses, the total number of electrons in a
fluid element (mesh cell) is $N_e=4\pi R^2dR\Gamma\rho/m_p$ where $dR$
is a cell width in the lab frame. The spectral power at the typical
frequency from a  cell is given by $F_{\nu_m} \propto N_e \Gamma
B$. Assuming a power-law distribution of the electron random Lorentz
factor with index $\hat{p}$, the observed flux is 
\begin{equation}
F_{\nu_m < \nu <
 \nu_c}=F_{\nu_m}\paren{\frac{\nu}{\nu_m}}^{-(\hat{p}-1)/2}
\propto N_e \Gamma^{(\hat{p}+1)/2}p^{(5\hat{p}-3)/4}\rho^{-(\hat{p}-1)}
\label{eq:fnu}
\end{equation}
where we have assumed that the observational band is located between
the typical frequency $\nu_m$ and the cooling frequency $\nu_c$. In
early afterglow phase, with the typical parameters, X-ray and optical band
satisfy this condition 
\footnote{
For the typical microscopic parameters $\epsilon_e=10^{-2}$ and
$\epsilon_B=10^{-3}$, the numerical results shown in this paper 
actually satisfy this condition in the time ranges of the plots 
( Figs \ref{thinLC} and \ref{ThickLC} ) where
$\epsilon_e$ and $\epsilon_B$ are the fractions of shock energy
given to magnetic field and electrons at a shock, respectively. 
Choosing a larger $\epsilon_B$ could produce an additional break at 
$t\sim 10^3-10^4$s especially in X-ray light curves due to the passage
of the forward-shock cooling frequency in X-ray band. 
However, the steepening is small $\Delta \alpha$=1/4 and it happens well
after the deceleration of the fireball. The cooling break does not
affect our discussions about the onset of afterglows.}
for the forward shock and reverse shock emission, respectively.
We evaluate the flux and observed time of photons from each shocked
fluid element, and superimpose the emission to construct the forward and
reverse shock light curves. The results are shown in figure
\ref{thinLC}. Although at later times the light curves approach the
expected power-law decays $L\propto t^{-1}$ (forward shock;
the thick solid line) and $L\propto t^{-2}$ (reverse shock; the thin
solid line), the peaks are rounded as we expected. 

At the deceleration time (and after that), the light crossing time of
the forward shocked region is comparable to the observed time $\Delta/c
\sim R/c\Gamma^2$. To evaluate the effect of the thickness of the forward
shocked region, we assume that all the shocked electrons emit photons at
the forward shock front. The resulting light curve peaks at an earlier
time 
\footnote{
A peak time is given by the sum of a deceleration time and a fraction of
the light crossing time of a shocked region.  If all the shock energy is
assumed to be radiated at the shock front, the emission should peak
earlier. The analytic deceleration time $t_d$ is obtained based on a
simple model, and it could have an error of a factor of a few.}
$t \sim t_d/2$, but the shape around the peak itself is similar. We
conclude that the round-off of the afterglow peak is mainly due to the
slow energy transfer from a fireball shell to a blast wave.

The internal shock model requires a highly irregular outflow from 
the GRB central engine. Since the hydrodynamic interaction inside 
the flow smooths the velocity and pressure profiles, but not the
density profile, fireball ejecta might have an irregular density 
profile at the deceleration time. Emission from the ejecta during a 
reverse shock crossing could reflect the light curve of the prompt 
emission produced by internal shocks (Nakar \& Piran 2004). 
The peak time of the reverse shock emission also depends on the density 
profile. As a fireball expands, a reverse shock accelerates more and
more electrons in the fireball while the Lorentz factor and pressure
of the shocked region decreases. The balance determines the peak time.
In the thin shell case, a reverse shock does not effectively decelerate 
the shell material that it crosses. This leads to  
large pressure gradient across the shocked region. The contribution 
of emission from the inner part of the shell becomes negligible, and 
the reverse shock emission peaks slightly before the shock crossing time 
(the thin solid line).

\subsection{The Thick Shell Case}
We consider the thick shell case in which a reverse shock becomes
relativistic during the propagation. The initial condition is
$E=10^{52}$ ergs, $\Gamma_0=10^3$, $R_0=3\times 10^{11}$cm, 
and $\rho_1=10$ $m_p$ cm$^{-3}$. The fireball is decelerated around
$R_d=l^{3/4}R_0^{1/4} \sim 1.5 \times 10^{16}$ cm, which corresponds to
the shock crossing time $t_d=R_0/c=10$ sec. 

\begin{figure}
\plotone{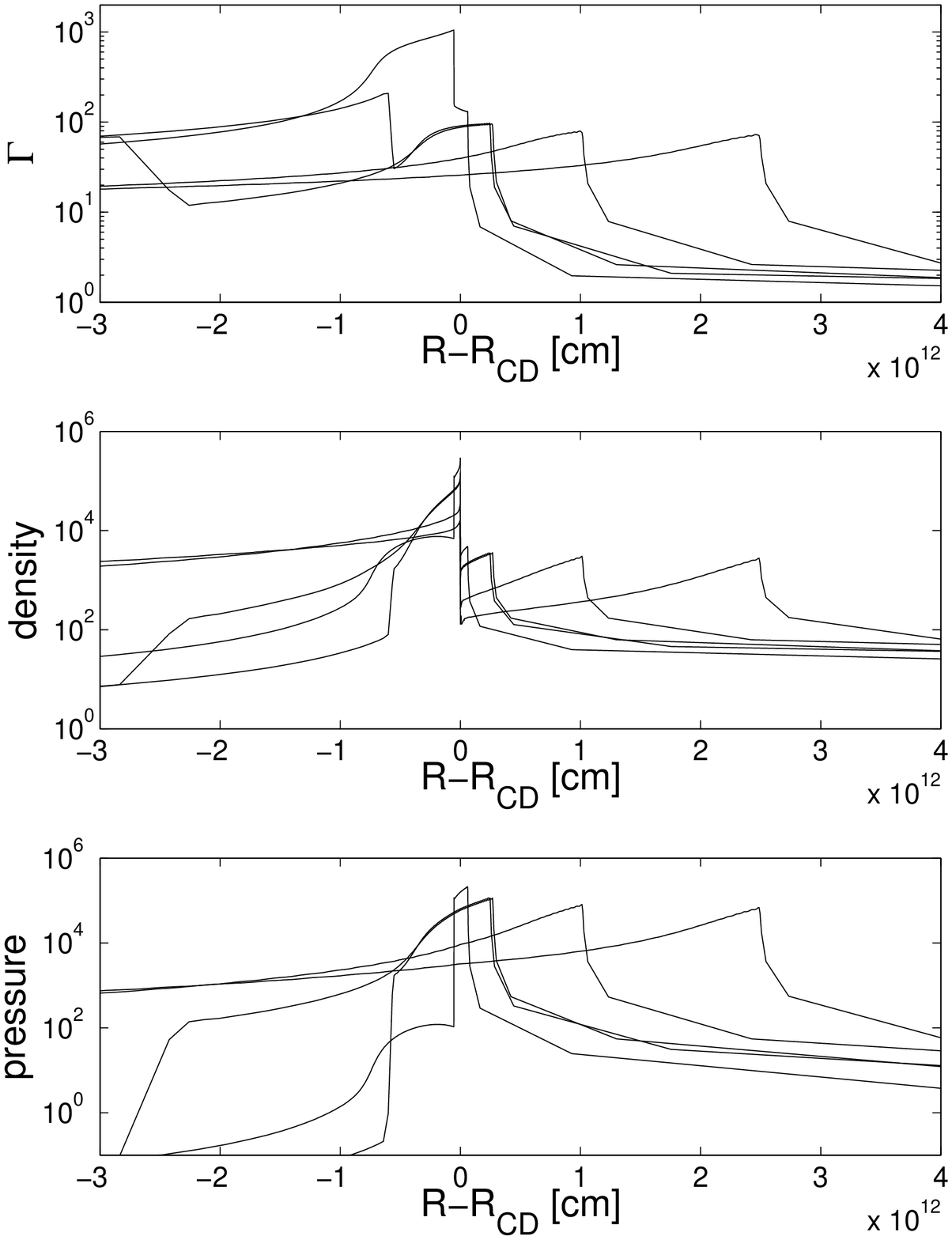}
\caption{Thick Shell Case: Profiles of $\Gamma$, $\rho$ and $p$ at
  different lab times 
  $\hat{t}=0.75 R_d/c$, $1.40 R_d/c$, $1.45R_d/c$, $1.92 R_d/c$ and
  $2.14R_d/c$. The $x-$axis is the distance from the
  contact discontinuity (CD). Photons emitted from the discontinuity
  at the lab times are observed at $t=5, 18, 19, 50$ and 
  100 s, respectively. The gamma-ray front
  is located at ($R-R_{CD})=ct$. 
 \label{profRRS}}
 \end{figure}

\begin{figure}
\plotone{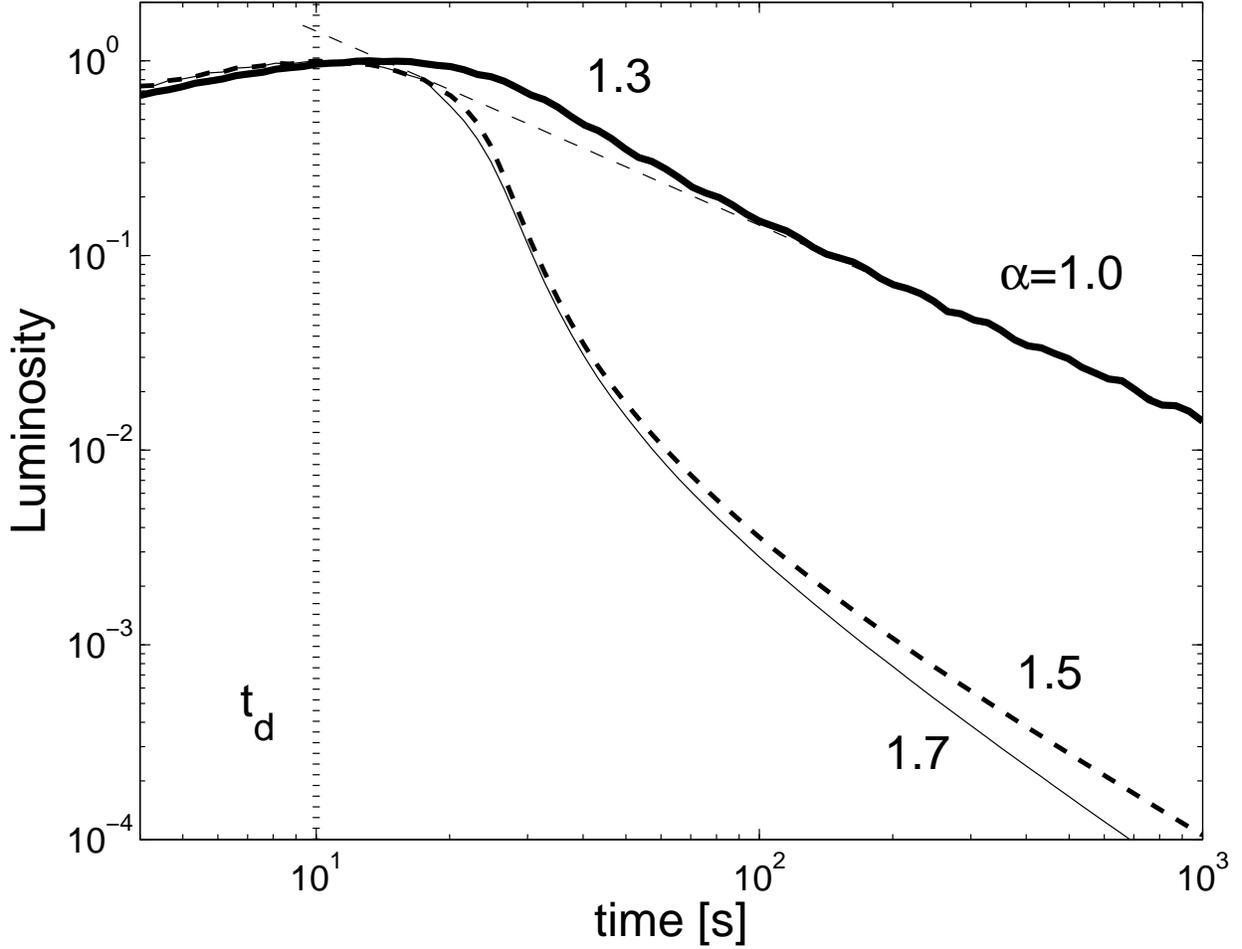}
\caption{Afterglow Light Curve: the Thick Shell Case.
$E=10^{52}$ ergs, $\Gamma_0=1000$, $R_0=3\times 10^{11}$cm, 
  $\rho_1=10$ $m_p$ cm$^{-3}$ and $\hat{p}=2.2$. 
  Forward shock emission in X-ray band (thick solid curve) 
  and reverse shock emission in optical band (thick dashed curve).
  Reverse shock emission from 
  the outer part of the shell which corresponds to 80\% of 
  the shell mass (thin solid curve). Forward shock emission 
  decays slightly faster at the beginning as $t^{-1.3}$ and later it
  decays as $t^{-1.0}$. reverse shock decay indexes $\alpha=1.5$  
  and $\alpha=1.7$. The deceleration time $t_d=R_0/c=10$ sec 
 (vertical dotted line). The light curves are normalized at
the peaks.
 \label{ThickLC}}
 \end{figure}

The energy transfer from the fireball to a blast wave is similar to 
that described in the thin shell case. At the shock crossing time, the
forward-shocked ISM and reverse-shocked shell have comparable energies
(the thick lines in figure \ref{energy}). The main difference is that 
in the thick shell case the shocked shell carries the energy in the form
of the internal energy, instead of the kinetic energy, because the
reverse shock significantly decelerate the shell. Figure \ref{profRRS}
depicts the profiles of $\Gamma$, $\rho$ and $p$. When the reverse shock
crosses the shell with a width of $\sim R_0$, a rarefaction wave begins
to propagate towards  
the contact discontinuity, and it quickly transfers the shell's internal
energy to the shocked ISM in a timescale comparable to 
the deceleration time (the shock crossing time). The steep decay of the 
shocked shell energy right after the peak (the thick dashed-dotted line) 
is due to the rarefaction wave propagation. Since for our parameters
the reverse shock is mildly relativistic in the frame of the unshocked shell 
$\bar{\Gamma} \sim \Gamma_0/(l/R_0)^{3/8} \sim $ several (Sari \& Piran
1995), the shocked shell becomes cold after the propagation,
and the energy of the shocked shell begins to decay in the same
rate as the thin shell does. 

In the thick shell case, the energy of a shocked shell is swiftly relayed
to a forward shock by a rarefaction wave. A broken power-law 
$\Gamma \propto R^{-1/2}$  $(R<R_d)$ and $\Gamma \propto R^{-3/2}$
$(R>R_d)$ describes the evolution of the forward shock well. The
round-off of an afterglow peak is expected to be small. 
Using eq. (\ref{eq:fnu}) and numerical results, we evaluate the light 
curves of the forward shock and reverse shock emission, which are shown
in figure \ref{ThickLC}.  Initially the luminosity of the both shocked
region slowly increases as $\sim t^{1/2}$ (Kobayashi 2000), and they
peak around the shock crossing time. The forward shock light curve (the
thick solid curve) is slightly steeper right after the peak compared to   
at late times. Although the decay is described well by a single power law
if the reference time $t_0$ is set at the middle between the GRB trigger
and the afterglow peak as we discussed in section 2, measuring
times from the GRB trigger also provides a reasonable estimate for the
afterglow decay. The reverse shock emission (the thick dashed curve)
drops sharply $\alpha \sim 5$ during the rarefaction wave
propagation. The high latitude (off-axis) emission could contribute to
the early flux (Kumar \& Panaitescu 2000). The numerical fireball shell
has a long tail in the density profile, and the shocked tail contributes
to the light curve at late times. The emission from the outer part
containing 80\% of the shell mass gives a steeper light curve (the thin
solid curve).

 \section{Inhomogeneous Fireball}
If a burst has a precursor, we might need a shift of the reference time
to a later time. A simple example is that a precursor with a negligible
energy is followed by the main component. The quiescent period could be
due to a dormant period in the central engine or due to the property of
the outflow (e.g. homogeneous outflow in which prompt gamma-rays are not 
produced by internal shocks). In the former case, $t_0$ should be set at
the explosion time of the main component. If the separation between the
precursor and the main component is much larger than the deceleration
time of the main component, measuring times from the precursor
introduces an artificial steeping in the early afterglow. 
Assuming a precursor and measuring times from it, we have replotted
figures \ref{thinLC} and \ref{ThickLC}. If the precursor is located  
at $\Delta t \sim 2000$sec $\sim 10 t_d$ or $\sim 3500$sec 
$\sim 18t_d$ before the main component, the decay index of the early
forward shock emission could be overestimated as $\alpha \sim 3$ or
$\sim 5$ in the thin shell case. In the thick shell case, $\Delta t=7
t_d$ and $12t_d$ lead to steep decay indexes of $\alpha\sim3$ and $\sim
5$, respectively. Since afterglow light curves are usually plotted with
$t_0$ setting at the GRB trigger time, precursors, although they might be
energetically small, need to be strong enough to trigger GRB detectors
(e.g. BAT) in order to cause an artificial steep decay.

\subsection{Case Studies}

X-ray flares were originally reported from BeppoSAX observations,
GRB 011121 and GRB 011211 (Piro et al 2005). 
GRB 011121 was the second-brightest GRB observed by BeppoSAX
in gamma-rays (after GRB 990123) and in X-rays (after GRB 010222). The
fluence in the 2-700 keV range corresponds to an isotropic energy of
$2.8 \times 10^{52}$ ergs at the redshift of the burst $z=0.36$. The
gamma-ray light curve shows a main peak starting at $t\sim 5$ sec and
ending at $t\sim30$ sec with minor substructures. An X-ray flare took
place at $t=240-310$ sec. The fluence of GRB 011211 in the 2-700 keV
range gives an isotropic energy of $3.6\times10^{52}$ ergs at the
redshift of the burst $z=2.14$. The gamma-ray prompt emission has a
long duration $T \sim 400$ sec, and  an X-ray  flare is detected from 600
to 700 sec. In the source frame, the two flares occurred at a similar time
$\sim 200$ sec and they have a width $\Delta/c\sim 30-50$ sec (see Piro
et al. 2005 and references therein for the basic parameters of these bursts). 

Since the decay part of the X-ray flares and the following shallower 
part are described with a single power-law when the time is measured
starting from the flare peak, the flares were suggested as
the beginning of the afterglow caused by  a thick shell (Piro et al. 2005).
The thick shell interpretation requires a long energy release from the
central engine. The burst should be long and extend all the way
to the  flare peak. However, the observed prompt emission ended 
well before the onset of the flares. In order to suggest the thick shell
case, one must assume that the efficiency of conversion into gamma-ray
varies dramatically with time. Furthermore, the engine should release
most fireball energy at the last moment (major reenergization at the
flare), because measuring times from the GRB trigger does not lead to an
overestimate of the decay index even in the thick shell case, as long as
the shell is homogeneous. 

In principle, the low conversion efficiency could originate from (1) a
small Lorentz factor of late ejecta which does not allow gamma-ray
radiation or (2) a small dispersion of the Lorentz factor of late ejecta
in the internal shock scenario. The critical Lorentz factor
is given by $\Gamma_c=(3E/4\pi\rho_1c^2\Delta^3)^{1/8} \sim 310
E_{53}^{1/8} \rho_{1,-1}^{-1/8} \Delta_{12}^{-3/8}$ 
where $Q_n=Q/10^n$ in the cgs unit and $\Delta_{12} \sim \Delta/(c\times 
40 sec)$. In the scenario (1), the low Lorentz factor condition 
$\Gamma \lesssim 100$ requires that the last energetic ejecta with 
an energy $E \gtrsim 10^{53}$ ergs and a width $\Delta/c \sim 40$ sec 
should be in the thin shell regime. The deceleration observed time 
$t_d \gtrsim 420 ~E_{53}^{1/3} \rho_{1,-1}^{-1/3}\Gamma_{2}^{-8/3}$ sec
becomes much larger than the width of the flares $\sim 40$ sec, and 
one finds that the scenario (1) is inconsistent
\footnote{ In the afterglow modeling of this thin shell case,
measuring times from the GRB trigger does not cause an early steep decay,
because the separation between the beginning of the prompt emission 
and the flare $\Delta t \sim 200$ sec is smaller than the deceleration
time $t_d \gtrsim 420$ sec. As we have shown at the beginning of this
section, a significant artificial steepening happens only when 
the ratio $\Delta t/t_d$ is larger than $\sim 10$.}.

To examine the scenario (2), we consider an inhomogeneous shell (two
components) expanding into ISM. Corresponding to the major energy release 
from the central engine at the last moment, the inner edge of the
shell with a width $\Delta_b/c \sim 40$ sec  is assumed to have an
energy $E_b$ larger than the energy $E_a \sim 10^{53}$ ergs in the
preceding outer part with a width $\Delta_a/c \sim 200$ sec. The Lorentz
factors of the two components at the end of the internal shock phase are
the same value of $\Gamma$. Both components should be in the thick shell
regime, otherwise the deceleration time of the shell becomes larger than
the flare occurrence time $\sim 200$ sec. The deceleration radius of the
shell should be larger than the radius $R \sim \Delta_a^{1/4}l_a^{3/4}$
at which a reverse shock crosses the outer component where
$l_a=l(E=E_a)$. The separation at the deceleration time between the
gamma-ray front and the reference BM blast wave satisfies $S_B \gtrsim 
(\Delta_a/8)(E_a/E_b)$. The separation at that time between the
gamma-ray front and the fireball is $S_F \sim \Delta_a+\Delta_b/4\sim
\Delta_a$ where we have assumed the fireball evolution $\Gamma \propto
R^{-1/2}$ before the deceleration. Note that a broke power law
description of $\Gamma$ around the deceleration radius is a good
approximation in the thick shell case. Then, we obtain $S_F/S_B \lesssim
8 (E_b/E_a)$. The precursor discussion at the beginning of this section
corresponds to the two  component model with $E_a=0$. We can show that
the separation ratio at the deceleration time is 
$S_F/S_B \sim 8(\Delta_a/\Delta_b)$ in this case. Since replotting 
figure \ref{ThickLC} we have found that $\Delta_a/\Delta_b \gtrsim 10$
to produce an early steep decay, in the scenario (2) $E_b$
should be at least ten times larger than $E_a$. This requires a large
energy budget for the central engine $E \gtrsim 10^{54}$ ergs. Even in the
limit of $E_a/E_b=0$, the small ratio $\Delta_a/\Delta_b \sim
200/40=5$ does not lead to a very steep decay of $\alpha \gtrsim 3-5$.
Therefore, we conclude that even an inhomogeneous fireball can not produce 
X-ray flares in early afterglow via external shock emission process.

 \section{Conclusions}

We have studied the reference time $t_0$ for the afterglow modeling.
Although measuring times from the beginning of the prompt emission (GRB
trigger) might cause a slight overestimate of the early afterglow slope
in the thick shell case. This choice of $t_0$ gives a reasonable
approximation, and it does not induce a very steep decay ($\alpha\sim
3-5$) like the early steep decay or X-ray flares  in the canonical Swift
X-ray light curve.  

The leading model to explain the rapid decay and flares in early X-ray 
afterglow is the internal shock emission. A clear, testable prediction
of this model is that the temporal decay index $\alpha$ of the tail part
should be related to the spectral index $\beta$ by an equation
$\alpha=2+\beta$ (Kumar \& Panaitescu 2000). 
When evaluating the emission decay $L\propto (t-t_0)^{-\alpha}$ in the 
internal shock model, an important difference is that the GRB trigger
time is no longer special. The reference time $t_0$ should correspond to
the onset of each particular spike in the prompt emission or in afterglow
(Kobayashi et al. 1997; Zhang et al. 2006a; Nousek et al. 2006; Fan \&
Wei 2005). Every time when the central engine is re-stared to eject
sub-shells, the reference time $t_0$ should be re-set to the
reactivation time of the engine.  
Although O'Brien et al. (2006) have found that $\alpha$ appears
to be largely independent of $\beta$ when the BAT trigger time is used
as $t_0$, Liang et al (2006) have shown that the relation
$\alpha=2+\beta$ is more or less satisfied in most cases if $t_0$ is set 
near the beginning of rising segment of the last pulse of the  
prompt emission or a corresponding X-ray flare, and if the underlying
forward shock emission component is subtracted. Swift observations
support the internal shock model.   

The self-consistent internal shock interpretation should be more
favorable than the beginning-of-the-afterglow interpretation. 
The latter can not explain multiple X-ray flares in a single event. Such
behavior is observed in many Swift bursts (Burrows et al. 2005; Falcone
et al. 2006; Romano et al. 2006; O'Brien et al 2006). Usually X-ray flux
already begins to decay before X-ray flares appear, and it suggests that
the onset of afterglow is prior to the flares. If a large amount of
energy is impulsively injected to a fireball during the deceleration,
$t_0$ might be re-set to the injection time. However, an afterglow
baseline also should shift after a flare (energy injection). This
clearly contradicts with observations in which after a flare peak,
afteglow decays back to its pre-flare flux level. We can not explain
X-ray flares by the shift of $t_0$ associated with large energy
injections. 

Swift discovered that a large fraction of X-ray afterglows have a slow
decay phase, and it is suggested that energy injection into a blast
wave takes place several hundred seconds after the burst. This implies
that right after the burst the kinetic energy of a blast wave is 
very low and in turn the efficiency of internal shock process is
extremely high (Zhang et al. 2006a; Nousek et al. 2006; Ioka et
al. 2005; Granot, K\"onigl \& Piran 2006; Zhang et al.2006b; however see 
also Fan \& Piran 2006). The round-off forward shock peak in the thin 
shell case might be a good candidate for the shallow decay phase.  
However, if we interpret the observed shallow decay as the round-shape
of the deceleration phase, the model light curve is shallower than some
of the observed ones. It may be because the observed curve is the
combination of this round phase and the rapid decay from the GRB tail
emission (high latitude). Equalizing the deceleration time and the
shallow phase timescale we obtain the initial Lorentz factor  
$\Gamma_0 \sim 110 \zeta^{3/8} E_{53}^{1/8} \rho_{1,-1}^{-1/8}t_{d,3}^{-3/8}$ 
where the time dilation effect is taken into account $\zeta=(1+z)/3$.

For a wind environment, we can discuss the $t_0$ issue in a very
similar way. The Lorentz factor of a shocked shell is constant in both
of the thin and thick shell cases during the shock crossing, while 
the BM blast wave decelerates as $\Gamma \propto R^{-1/2}$ (Kobayashi \&
Zhang 2003). The separations are comparable $S_F \sim 2S_B$ at the
deceleration time. Measuring times from the beginning of the prompt
phase should be a good approximation for events in a wind environment also.
Since most bursts in a wind environment fall in the thick shell case 
(Kobayashi, \Mesz \& Zhang 2004), and since the deceleration time is 
comparable to GRB duration in the thick shell case, another process 
(e.g. refreshed shocks) rather than the afterglow peak is necessary to 
explain the shallow decay phase. 

 We thank Luigi Piro for useful discussion. This work is supported by
 NASA NNG05GB67G and NNG06GH62G.

\end{document}